\documentclass[twocolumn]{aastex61}
\pdfoutput=1 
\usepackage{amsmath,amstext}
\usepackage[T1]{fontenc}
\usepackage{apjfonts} 
\usepackage[figure,figure*]{hypcap}

\newcommand{\HII}{H\,\textsc{ii}}
\newcommand{\HI}{H\,\textsc{i}}

\newcommand{\NII}{N\,\textsc{ii}}
\newcommand{\OII}{O\,\textsc{ii}}
\newcommand{\OIII}{O\,\textsc{iii}}
\newcommand{\SII}{S\,\textsc{ii}}

\shorttitle{The Little Cub: a near-pristine galaxy}
\shortauthors{Hsyu et al.}

\begin{document}

\title{The Little Cub: discovery of an extremely metal-poor star-forming galaxy in the local universe}

\author{Tiffany Hsyu}
\affiliation{Department of Astronomy \& Astrophysics, University of California Santa Cruz, 1156 High Street, Santa Cruz, CA 95060, USA}
\author{Ryan J. Cooke}
\affiliation{Centre for Extragalactic Astronomy, Department of Physics, Durham University, South Road, Durham DH1 3LE, UK}
\author{J. Xavier Prochaska}
\affiliation{Department of Astronomy \& Astrophysics, University of California Santa Cruz, 1156 High Street, Santa Cruz, CA 95060, USA}
\author{Michael Bolte}
\affiliation{Department of Astronomy \& Astrophysics, University of California Santa Cruz, 1156 High Street, Santa Cruz, CA 95060, USA}

\begin{abstract}
We report the discovery of the Little Cub, an extremely metal-poor star-forming galaxy in the local universe, found in the constellation Ursa Major (a.k.a. the Great Bear). We first identified the Little Cub as a candidate metal-poor galaxy based on its Sloan Digital Sky Survey photometric colors, combined with spectroscopy using the Kast spectrograph on the Shane 3 m telescope at Lick Observatory. In this Letter, we present high-quality spectroscopic data taken with the Low Resolution Imaging Spectrometer at Keck Observatory, which confirm the extremely metal-poor nature of this galaxy. Based on the weak [\OIII] $\lambda$4363\,\AA\,emission line, we estimate a direct oxygen abundance of 12\,+\,log(O/H)\,=\,7.13\,$\pm$\,0.08, making the Little Cub one of the lowest-metallicity star-forming galaxies currently known in the local universe. The Little Cub appears to be a companion of the spiral galaxy NGC 3359 and shows evidence of gas stripping.  We may therefore be witnessing the quenching of a near-pristine galaxy as it makes its first passage about a Milky Way--like galaxy. 

\end{abstract}


\section{Introduction}

The observed galaxy luminosity function (LF; \citealt{1976ApJ...203..297S}) indicates that low-luminosity, low-mass galaxies are the most common type of galaxy in the universe. From the luminosity--metallicity (L--Z) relation \citep{1989ApJ...347..875S, 2001A&A...374..412P, 2012ApJ...754...98B}, we then expect these low-luminosity galaxies to be among the least chemically evolved environments in the universe. 

Metal-poor environments provide a unique opportunity to study the conditions under which the first stars might have formed in the early universe and the subsequent chemical evolution of galaxies (e.g., \citealt{2017MNRAS.467..802C}). Local studies of metal-poor systems have typically focused on nearby blue compact dwarf (BCD) galaxies, whose active or recent star formation produces observable emission lines from their \HII\ regions. BCDs are also of particular interest for determining the primordial helium abundance that was set during Big Bang Nucleosynthesis \citep{2014MNRAS.445..778I, 2015JCAP...07..011A}, constraining  the physical properties of massive, metal-poor stars \citep{2005ApJ...621..269T} that dominated the chemical and physical evolution of the first galaxies, and understanding how these stellar populations interacted with and enriched their surroundings \citep{2001ApJ...555...92M, 2003ApJ...588...18F}. 


However, the number of observed low-luminosity, low-metallicity systems is much smaller than the number of systems predicted by the LF \citep{2011ApJ...743...77M}. Although the detection of low-luminosity systems presents an observational challenge due to their intrinsically low surface brightness, observational biases alone cannot account for the dearth of observed BCDs \citep{2017ApJ...835..159S}. 
Numerous efforts have focused on the detection of new BCDs \citep{2012A&A...546A.122I, 2015MNRAS.448.2687J, 2016arXiv161106881G, 2017arXiv170101011G, 2017MNRAS.465.3977J}, but progress has been slow, particularly in the \emph{lowest}-metallicity regime. Aside from the well-known, low-metallicity systems I Zwicky 18 \citep{1966ApJ...143..192Z}, SBS\,0335$-$052 \citep{1990Natur.343..238I}, and DDO68 \citep{2005A&A...443...91P},  which exhibit higher luminosities given their metallicity \citep{2010MNRAS.406.1238E},
only two new systems that push on the lowest-metallicity regime of the L--Z relation have been discovered: Leo P \citep{2013AJ....146....3S} and AGC\,198691 \citep{2016ApJ...822..108H}, both through the blind \HI\ 21~cm line Arecibo Legacy Fast ALFA survey (ALFALFA; \citealt{2005AJ....130.2598G, 2011AJ....142..170H}). 

In this Letter, we report the discovery of one of the lowest-metallicity BCDs currently known, J1044+6306, found in the constellation Ursa Major (a.k.a. the Great Bear), which we nickname the Little Cub. In Section  \ref{ODR}, we present the results of our spectroscopic observations of the Little Cub made using the Kast spectrometer on the Shane 3 m telescope at Lick Observatory and the Low Resolution Imaging Spectrometer (LRIS)  at Keck Observatory.  We analyze the spectra and derive chemical abundances in Section \ref{AD}. In Section \ref{Disc}, we discuss the physical properties of the Little Cub and the environment in which it resides, including its potential interaction with the nearby spiral galaxy NGC 3359. We conclude our findings in Section \ref{Conc}.

\section{Observations and Data Reduction}
\label{ODR}
The Little Cub was selected as a candidate BCD based on its photometric colors in the Sloan Digital Sky Survey (SDSS) Data Release 12 (DR12). Its discovery is part of a larger program led by the authors to increase the current meager population of the lowest-metallicity galaxies using photometry alone, thus circumventing the need for pre-existing spectroscopic information. Specifically, we queried SDSS DR12 for extended objects (i.e., classified as a galaxy by SDSS) with photometric colors similar to those of Leo P, I Zwicky 18, and other known low metallicity systems.  We visually examined the SDSS imaging of 2505 candidate BCDs to eliminate contaminants, such as star-forming \HII\ regions in larger galaxies, and selected a subset of systems to observe and estimate their metallicity. Currently, follow-up spectroscopy has been obtained for 158 candidate BCDs; of these, about 100 new BCDs have been identified, including the Little Cub. Further details of this survey, including the full sample of systems discovered so far, will be presented in a forthcoming paper (T. Hsyu et al. 2017, in preparation).



\subsection{Lick Observations}
The Little Cub was first observed on 2016 February 2 using the Kast spectrograph on the Shane 3 m telescope at Lick Observatory. The Kast spectrograph is equipped with separate blue and red channels.  We used the 600/4310 grism on the blue side, the 1200/5000 grating on the red side, and the d55 dichroic, for an approximate wavelength coverage of  3300--5500\,\AA\,and  5800--7300\,\AA.

Observations were made using a 2$''$ slit aligned at parallactic angle to best compensate for differential atmospheric refraction, for a total exposure time of 3$\times$1800s on both the red and blue channels. The spectrophotometric standard star Feige 66 was observed to calibrate the flux of our spectra. Exposures of the Hg--Cd and He arc lamps on the blue side and the Ne arc lamp on the red side were obtained at the beginning of each night for wavelength calibration. Bias frames and dome flats were also obtained to correct for the detector bias level and pixel-to-pixel sensitivity variations.\footnote{The data presented in this Letter were reduced with the \textsc{pypit} spectroscopic data reduction pipeline, which is available from: \url{https://github.com/PYPIT/PYPIT}}

Initial observations of the Little Cub included the detection of the [\OII] doublet at $\lambda\lambda$3727,3729\,\AA $ $,  H$\beta$ emission at $\lambda$4861\,\AA $ $, the [\OIII] doublet at $\lambda\lambda$4959,5007\,\AA $ $, H$\alpha$ emission at $\lambda$6563\,\AA $ $, the [\NII] doublet at $\lambda\lambda$6548,6584\,\AA $ $, and the [\SII] doublet at $\lambda\lambda$6717,6731\,\AA $ $. The temperature-sensitive oxygen line at [\OIII] $\lambda$4363\,\AA\, necessary for a direct oxygen abundance measurement is not detected in our Kast observations. To obtain a first guess of the metallicity, we 
assumed an electron density of $n_{e}$\,=\,100 cm$^{-3}$ and an electron temperature of $T_{e}$\,=\,17,000 K in the high ionization zone, which are values typical of \HII\,regions (similar values were derived for the \HII\,region in Leo P; \citealt{2013AJ....146....3S}). This method indicated that the Little Cub is extremely metal-poor, with an estimated metallicity of 12\,+\,log(O/H) \,$\lesssim$\,7.26.

\subsection{Keck Observations} Follow-up observations of the Little Cub were made using LRIS at Keck Observatory on 2016 February 16 and 2016 April 3, with the goal of detecting the [\OIII] $\lambda$4363\,\AA\,line to obtain a direct oxygen abundance measurement. Using the D560 dichroic, the 600/4000 grism on the blue side, and the 600/7500 grating on the red side, we achieved a total wavelength coverage of 3200--8600\,\AA, including a $\sim$200\,\AA\,overlap between the red and blue spectra near the dichroic ($\sim$5600\AA). We acquired 3$\times$600s and 6$\times$300s exposures in February on the red and blue sides respectively, and in April we acquired 3$\times$1200s and 2$\times$1875s exposures on the red and blue sides. The total exposure time on the red and blue arms are 5400s and 5550s, respectively.

\begin{figure}
\centering
\includegraphics[width=0.5\textwidth]{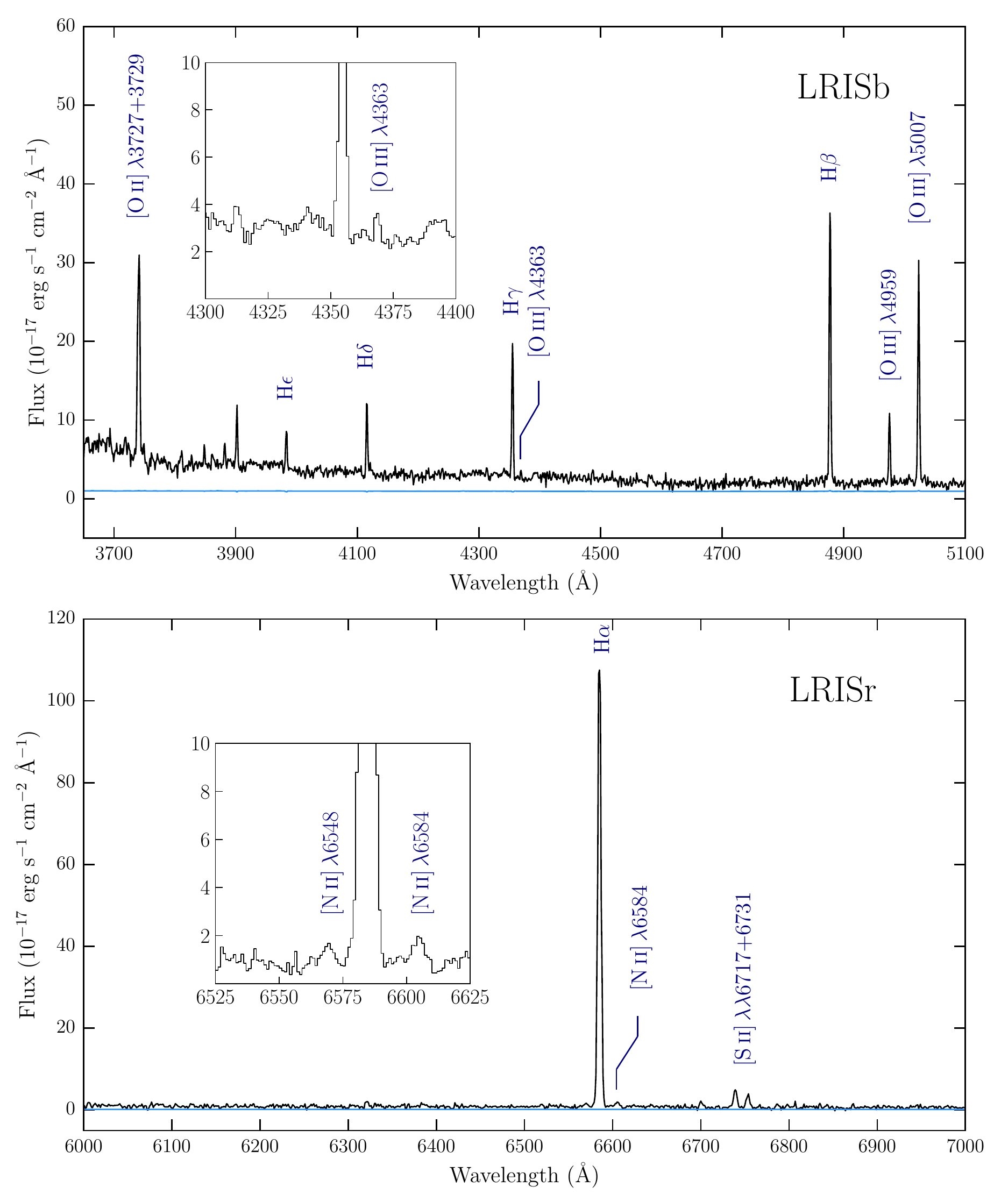}
\caption{Emission line spectra of the Little Cub (black histogram) obtained using LRIS at Keck Observatory. The corresponding error spectrum is shown in blue. The upper and lower panels represent the data collected using the separate blue and red channels, respectively. The inset in the upper panel shows a zoom-in of the temperature-sensitive [\OIII]\,$\lambda$\,4363\,\AA\,line, which is necessary for a direct oxygen abundance measurement. The inset in the lower panel shows a zoom-in of the weak [\NII]\,$\lambda$\,6584\,\AA\,line. A handful of emission lines that are used in our analysis are labelled.}
\label{fig:SDSSspec}
\end{figure}

We used a $0.7^{\prime\prime}$ slit and observed in $0.6^{\prime\prime}$ and $0.8^{\prime\prime}$ seeing in February and April, respectively, and used the atmospheric dispersion corrector. Bias frames and dome flats were obtained at the beginning of the night, along with spectra of the Hg, Cd, and Zn arc lamps on the blue side and Ne, Ar, and Kr arc lamps on the red side for wavelength calibration. The spectrophotometric standard star Feige 66 was observed in February and HZ 44 in April for flux calibration.
The spectra that we report here are extracted using a boxcar kernel of width 6.2$''$. The one-dimensional reduced, combined, and flux-calibrated spectrum of the Little Cub taken with Keck+LRIS is shown in Figure~\ref{fig:SDSSspec}.\footnote{The individual exposures were combined using \textsc{UVES$\_$popler}, which can be found at: \url{http://astronomy.swin.edu.au/~mmurphy/UVES_popler/}}

\section{Analysis and Discussion}
\label{AD}
\subsection{Emission Line Measurements}
Emission line fluxes were measured using the Absorption LIne Software (ALIS), which performs spectral line fitting using chi-squared minimization.\footnote{\textsc{ALIS} is available from: \url{https://github.com/rcooke-ast/ALIS/}} We model the intrinsic shape of each emission line with a Gaussian, where the full width at half maximum (FWHM) of the intrinsic profile is allowed to vary during the least-squares minimization. We convolve this model with the instrument line spread function, assumed to be a Gaussian. Based on the measured widths of the sky emission lines, we determine the instrument FWHM resolution to be 2.6\,\AA\ and 3.1\,\AA\ on the blue and red channels, respectively, which we use throughout our analysis. The continuum level around each emission line is fit simultaneously with the Gaussian, assuming a first-order Legendre polynomial; any uncertainty in the continuum placement is therefore folded into the uncertainty on our measured fluxes. All emission lines are tied to have the same intrinsic FWHM and we note that this assumption is justified, as the intrinsic width of the emission lines is much smaller than the instrumental broadening.

As a sanity check of the above modeling, we also measured the integrated flux above the continuum level of each emission line. These values, together with the Gaussian model values, are listed in the first three columns of Table \ref{table:fluxes}. We find that the resulting flux values from the separate methods, the derived physical properties, and the metallicity, are in good agreement. Henceforth, we adopt the values based on our Gaussian modeling procedure.

\begin{deluxetable*}{cccc|ccc}[ht!]
\tablecaption{Emission line Fluxes, Intensities, and Physical Properties of the Little Cub}
\tablewidth{0pt}
\tablehead{
\colhead{} & \colhead{} & $F(\lambda)/F(\mbox{H}\beta)$ & \colhead{} & \multicolumn{3}{c}{$I(\lambda)/I(\mbox{H}\beta)$} }
\startdata
{Ion} & {Shane/Kast} & {Keck/LRIS} & {Keck/LRIS} & {Shane/Kast} & {Keck/LRIS} & {Keck/LRIS} \\
{} & {} & {(Gaussian Model)} & {(Integrated Flux)} & {} & {(Gaussian Model)} & {(Integrated Flux)} \\
\hline
{[O\,\textsc{ii}]\,$\lambda$3727+3729} & 1.187\,$\pm$\,0.042 & 1.017\,$\pm$\,0.014 & 1.049\,$\pm$\,0.015 & 1.361\,$\pm$\,0.042 & 1.056\,$\pm$\,0.021 & 1.071\,$\pm$\,0.021 \\
{H11\,$\lambda$3771} & \nodata & 0.0192\,$\pm$\,0.0051 & 0.0181\,$\pm$\,0.0060 & \nodata & 0.0199\,$\pm$\,0.0053 & 0.0185\,$\pm$\,0.0061 \\
{H10\,$\lambda$3798} & \nodata & 0.0442\,$\pm$\,0.0055 & 0.0499\,$\pm$\,0.0059 & \nodata & 0.0458\,$\pm$\,0.0058 & 0.0509\,$\pm$\,0.0061 \\
{H9\,$\lambda$3835} & \nodata & 0.0511\,$\pm$\,0.0046 & 0.0532\,$\pm$\,0.0053 & \nodata & 0.0529\,$\pm$\,0.0048 & 0.0542\,$\pm$\,0.0055 \\
{[Ne\,\textsc{iii}]\,$\lambda$3868} & 0.094\,$\pm$\,0.023 & 0.0620\,$\pm$\,0.0057 & 0.0631\,$\pm$\,0.0064 & 0.108\,$\pm$\,0.023 & 0.0636\,$\pm$\,0.0060 & 0.0643\,$\pm$\,0.0065 \\
{H8+He\,\textsc{i}\,$\lambda$3889} & 0.173\,$\pm$\,0.021 & 0.1702\,$\pm$\,0.0047 & 0.1763\,$\pm$\,0.0071 & 0.198\,$\pm$\,0.021 & 0.1759\,$\pm$\,0.0053 & 0.1795\,$\pm$\,0.0075 \\
{H$\epsilon$+[Ne\,\textsc{iii}]\,$\lambda$3968} & 0.168\,$\pm$\,0.020 & 0.1182\,$\pm$\,0.0020 & 0.1218\,$\pm$\,0.0028 & 0.192\,$\pm$\,0.019 & 0.1219\,$\pm$\,0.0025 & 0.1238\,$\pm$\,0.0032 \\
{H$\delta$\,$\lambda$4101} & 0.220\,$\pm$\,0.018 & 0.2441\,$\pm$\,0.0062 & 0.2381\,$\pm$\,0.0069 & 0.252\,$\pm$\,0.018 & 0.2505\,$\pm$\,0.0068 & 0.2415\,$\pm$\,0.0074 \\
{H$\gamma$\,$\lambda$4340} & 0.585\,$\pm$\,0.027 & 0.4499\,$\pm$\,0.0063 & 0.4595\,$\pm$\,0.0064 & 0.670\,$\pm$\,0.028 & 0.4586\,$\pm$\,0.0072 & 0.4644\,$\pm$\,0.0073 \\
{[O\,\textsc{iii}]\,$\lambda$4363} & \nodata & 0.0221\,$\pm$\,0.0046 & 0.0259\,$\pm$\,0.0051 & \nodata & 0.0225\,$\pm$\,0.0047 & 0.0261\,$\pm$\,0.0052 \\
{He\,\textsc{i}\,$\lambda$4472} & \nodata & 0.0235\,$\pm$\,0.0036 & 0.0281\,$\pm$\,0.0034 & \nodata & 0.0239\,$\pm$\,0.0036 & 0.0283\,$\pm$\,0.0035 \\
{H$\beta$\,$\lambda$4861} & 1.000\,$\pm$\,0.018 & 1.000\,$\pm$\,0.016 & 1.000\,$\pm$\,0.035 & 1.000\,$\pm$\,0.018 & 1.000\,$\pm$\,0.016 & 1.000\,$\pm$\,0.035 \\
{He\,\textsc{i}\,$\lambda$4922} & \nodata & 0.0116\,$\pm$\,0.0040 & 0.0127\,$\pm$\,0.0031 & \nodata & 0.0116\,$\pm$\,0.0040 & 0.0127\,$\pm$\,0.0031 \\
{[O\,\textsc{iii}]\,$\lambda$4959} & 0.240\,$\pm$\,0.017 & 0.2371\,$\pm$\,0.0041 & 0.2263\,$\pm$\,0.0046 & 0.240\,$\pm$\,0.017 & 0.2363\,$\pm$\,0.0041 & 0.2259\,$\pm$\,0.0046 \\
{[O\,\textsc{iii}]\,$\lambda$5007} & 0.756\,$\pm$\,0.016 & 0.7456\,$\pm$\,0.0058 & 0.7661\,$\pm$\,0.0067 & 0.756\,$\pm$\,0.016 & 0.7418\,$\pm$\,0.0060 & 0.7639\,$\pm$\,0.0068 \\
{He\,\textsc{i}\,$\lambda$5015}  & \nodata & 0.0230\,$\pm$\,0.0050 & 0.0230\,$\pm$\,0.0047 & \nodata & 0.0229\,$\pm$\,0.0050 & 0.0229\,$\pm$\,0.0047 \\
{He\,\textsc{i}\,$\lambda$5876} & \nodata & 0.0815\,$\pm$\,0.0066 & 0.0798\,$\pm$\,0.0064 & \nodata & 0.0807\,$\pm$\,0.0059 & 0.0784\,$\pm$\,0.0064 \\
{[N\,\textsc{ii}]\,$\lambda$6548} & \nodata & 0.0196\,$\pm$\,0.0028 & 0.0192\,$\pm$\,0.0028 & \nodata & 0.0209\,$\pm$\,0.0027 & 0.0187\,$\pm$\,0.0027 \\
{H$\alpha$\,$\lambda$6563} & 3.165\,$\pm$\,0.018 & 2.883\,$\pm$\,0.011 & 2.824\,$\pm$\,0.011 & 2.750\,$\pm$\,0.018 & 2.750\,$\pm$\,0.010 & 2.750\,$\pm$\,0.011 \\
{[N\,\textsc{ii}]\,$\lambda$6584} & \nodata & 0.0304\,$\pm$\,0.0028 & 0.0298\,$\pm$\,0.0027 & \nodata & 0.0272\,$\pm$\,0.0025 & 0.0290\,$\pm$\,0.0027 \\
{He\,\textsc{i}\,$\lambda$6678} & \nodata & 0.0363\,$\pm$\,0.0032 & 0.0355\,$\pm$\,0.0031 & \nodata & 0.0344\,$\pm$\,0.0033 & 0.0345\,$\pm$\,0.0031 \\
{[S\,\textsc{ii}]\,$\lambda$6717} & 0.1073\,$\pm$\,0.0078 & 0.1154\,$\pm$\,0.0022 & 0.1130\,$\pm$\,0.0021 & 0.0933\,$\pm$\,0.0078 & 0.1122\,$\pm$\,0.0028 & 0.1099\,$\pm$\,0.0029 \\
{[S\,\textsc{ii}]\,$\lambda$6731} & 0.0832\,$\pm$\,0.0075 & 0.0806\,$\pm$\,0.0021 & 0.0789\,$\pm$\,0.0020 & 0.0723\,$\pm$\,0.0075 & 0.0776\,$\pm$\,0.0024 & 0.0767\,$\pm$\,0.0024 \\
{He\,\textsc{i}\,$\lambda$7065} & \nodata & 0.0224\,$\pm$\,0.0017 & 0.0220\,$\pm$\,0.0016 & \nodata & 0.0262\,$\pm$\,0.0018 & 0.0213\,$\pm$\,0.0016 \\
\hline
c(H$\beta$) & & & & 0.00\,$\pm$\,0.10 & 0.06\,$\pm$\,0.02 & 0.04\,$\pm$\,0.02\\
F(H$\beta$) ($\times$10$^{-17}$ erg\,s$^{-1}$\,cm$^{-2}$) & & & & 1580.6\,$\pm$\,9.1 & 484.4\,$\pm$\,1.8 & 494.7\,$\pm$\,1.9 \\
EW(H$\beta$) (\AA) & & & & 101.6\,$\pm$\,8.2 & 53.6\,$\pm$\,1.6 & 50.6\,$\pm$\,3.8 \\
\hline \hline
Derived Physical Properties & & & & & Value & \\
\hline
$n_{e}$(\,[\SII]\,) (cm$^{-3}$) & & & & 180\,$^{+180}_{-110}$ & 32\,$^{+34}_{-17}$ & 39\,$^{+38}_{-23}$ \\
$T_{e}$(\,[\OIII]\,) (K) & & & & 17000 & 18700\,$\pm$\,2300 & 20100\,$\pm$\,2500 \\
$T_{e}$(\,[\OII]\,) (K) & & & & 14500 & 15000\,$\pm$\,800 & 15400\,$\pm$\,790 \\
O$^{+}$/H$^{+}$ ($\times$10$^{6}$) & & & & 12.14\,$\pm$\,0.48 & 8.61\,$\pm$\,1.62 & 8.02$\pm$\,1.38 \\
O$^{++}$/H$^{+}$ ($\times$10$^{6}$) & & & & 6.12\,$\pm$\,0.18 & 5.17\,$\pm$\,2.39 & 4.50\,$\pm$\,1.34 \\
12\,+\,log(O/H) & & & & 7.26\,$\pm$\,0.01 & 7.13\,$\pm$\,0.08 & 7.09\,$\pm$\,0.08\\
\enddata
\tablecomments{All calculations of electron temperature and ionic abundances assume an electron density of $n_{e}$\,=\,100\,cm$^{-3}$. For measurements based on our Kast data, where we do not detect the temperature-sensitive [\OIII]\,$\lambda$4363\,\AA\,line, we assume $T_{e}$\,=\,17,000\,K in the high ionization zone and $T_{e}$\,=\,14,500\,K in the low ionization zone, which is typical of metal-poor \HII\,regions \citep{2013AJ....146....3S}.}
\label{table:fluxes}
\end{deluxetable*}

The measured emission line fluxes are corrected for reddening and underlying stellar absorption simultaneously using the $\chi^{2}$ minimization approach described in \citet{b9890a982907415fa4d3eb85829e9388}, using our observed H$\beta$, H$\gamma$, and H$\delta$ fluxes.  We find that the underlying stellar absorption is $\lesssim$\,1\,\AA. The total reddening is minimal, with \textit{c}(H$\beta$)\,$\simeq$\,0.05, or $A_{V}$\,$\simeq$\,0.1 magnitudes.  Assuming a foreground extinction value of $A_{V}$\,=\,0.019 magnitudes from the \citet{2011ApJ...737..103S} Galactic dust reddening map implies only a small amount of internal reddening in the Little Cub, which is consistent with its low metallicity.

The emission line ratios are corrected for the uncertainty of the relative flux calibration across the separate blue and red channels by scaling our measurements of H$\alpha$ and H$\beta$ to the theoretical Balmer line ratios of an \HII\ region of electron temperature $T_{e}$\,=\,19,000~K.
We note that the scaling of fluxes resulting from these separate channels does not affect the results of our oxygen abundance measurement of the Little Cub. The corrected emission line ratios, normalized to the measured H$\beta$ flux, are presented in the final three columns of Table \ref{table:fluxes} for the Shane+Kast and the Keck+LRIS observations. 

\subsection{Metallicity Measurement}
Calculations of the electron density ($n_{e}$) and electron temperature ($T_{e}$) were performed using \textsc{PyNeb} \citep{2015A&A...573A..42L}; the results of these calculations are presented in the bottom section of Table \ref{table:fluxes}.\footnote{\textsc{PyNeb} is available from: \url{http://www.iac.es/proyecto/PyNeb/}} 

We use the [\SII] $\lambda\lambda$6717,6731\,\AA\,doublet, which is significantly detected in both the Kast and LRIS data, to estimate the electron density. These data show that the Little Cub's \HII\ region is in the low-density regime, where the [\SII] doublet is less sensitive to density \citep{1989agna.book.....O}. Given that our data only afford an upper limit on the electron density, we assume a value of $n_{e}$ = 100~cm$^{-3}$ in the subsequent ionic abundance measurements, which is both consistent with the expected low-density regime and with the density as determined by the [\SII] $\lambda\lambda$6717,6731\,\AA\,lines.\footnote{We note that adopting the 2$\sigma$ upper limit on the electron density yields a metallicity that agrees with our reported value.}

For electron temperature measurements, we assume a two-zone approximation of the \HII\ region and calculate separate electron temperatures corresponding to the high and low ionization regions. The electron temperature of the high ionization zone is measured using the ratio of the [\OIII] $\lambda$4363\,\AA\,and [\OIII] $\lambda$5007\,\AA\,lines. Based on our LRIS data, we estimate the electron temperature to be $T_{e}$\,=\,18700\,$\pm$\,2300~K. We do not detect the [\OII] $\lambda\lambda$7320,7330\,\AA\,or the [\NII] $\lambda$5755\,\AA\,lines, which are necessary for a direct measurement of the temperature in the low ionization zone. We therefore estimate a temperature using the \citet{1992MNRAS.255..325P} relation between the high and low ionization zone temperatures, and we include a systematic uncertainty of $\pm$500\,K on our estimate of the low ionization zone temperature to account for the spread of model values used to construct the \citet{1992MNRAS.255..325P} relation. The high and low ionization electron temperatures combined with the assumed electron density provide a measure of the gas-phase oxygen abundance, 12\,+\,log(O/H)\,=\,7.13\,$\pm$\,0.08.

\begin{figure}
\centering
\includegraphics[width=0.5\textwidth]{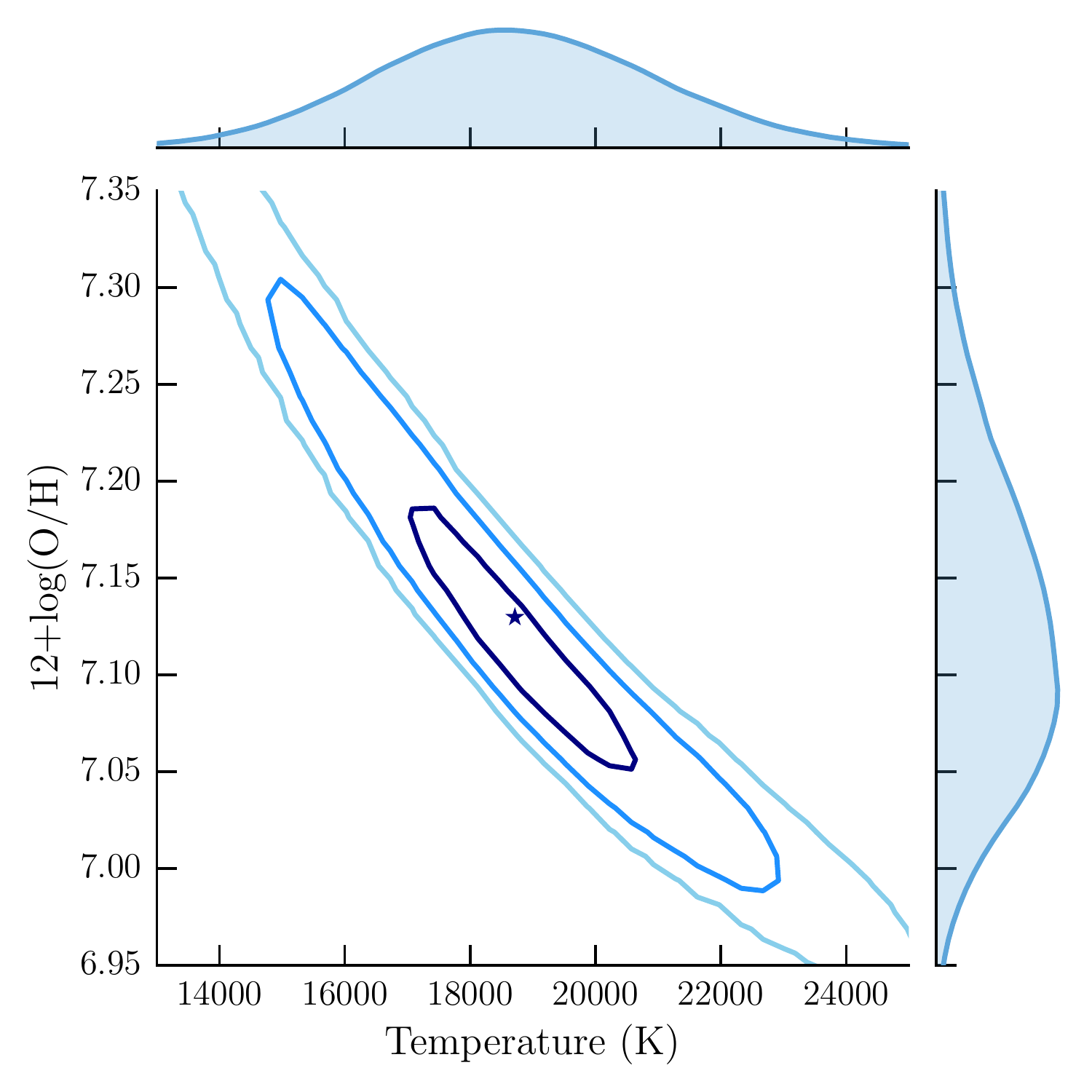}
\caption{Results of $10^{6}$ Monte Carlo realizations of the electron temperature in the high ionization zone and the resulting oxygen abundance. The contours represent the 1$\sigma$, 2$\sigma$, and 3$\sigma$ levels, and the starred symbol represents the most likely value of the temperature and metallicity. The margins show the projected distribution of the temperature and metallicity.}
\label{fig:contour}
\end{figure}

The dominant uncertainty of this metallicity measurement is the electron temperature, specifically the emission line flux of the [\OIII] $\lambda$4363\,\AA\,line. Overestimating the  [\OIII] $\lambda$4363\,\AA\,flux yields an inflated temperature measurement, which results in a lower oxygen abundance. Given that the [\OIII] $\lambda$4363\,\AA\,line is weak, we have specifically designed our observations to obtain a confident measure of its integrated flux, and here we report a $\gtrsim$5$\,\sigma$ detection. To illustrate the sensitivity of our measurement to the inferred oxygen abundance, we perturbed the [\OII], [\OIII], and H$\beta$ line fluxes by their measurement errors to construct $10^{6}$ Monte Carlo realizations. We then calculated the resulting distributions of electron temperature, ionic abundances, and metallicity of each realization. Our quoted temperature and metallicity are based on the mean of these calculations, which are presented in Figure \ref{fig:contour}.

\section{Discussion}
\label{Disc}
\subsection{Distance and Properties}
There are currently no reliable distance measurements to the Little Cub. We present two separate distance estimates using the \citet{2000ApJ...529..786M} and \citet{2005PhDT.........2M} flow models, which correct for the local velocity field. These models predict the distance to the Little Cub to be 20.6~Mpc and 15.8~Mpc, respectively. We note that flow model estimates can be highly uncertain for nearby galaxies. Calculations of distance-dependent properties are listed in Table \ref{table:properties} and described below.

The H$\alpha$ luminosity, \textit{L}(H$\alpha$), is determined using our measured H$\alpha$ flux combined with our assumed distances. The star formation rate (SFR) is derived using the relation between \textit{L}(H$\alpha$) and SFR \citep{1998ARA&A..36..189K}. We have divided the \citet{1998ARA&A..36..189K} SFR by a factor of 1.8; this correction accounts for the flattening of the stellar initial mass function (IMF) below $1~M_{\odot}$ \citep{2003PASP..115..763C} relative to the power-law Salpeter IMF used by \citet{1998ARA&A..36..189K}. There is some additional uncertainty in this conversion from \textit{L}(H$\alpha$) to a SFR due to the metal-poor nature of the Little Cub; O stars may be more efficient ionizers in low metallicity environments than their counterparts in more metal-rich environments, from which the \citet{1998ARA&A..36..189K} calibration is derived. This may lead to an overestimate of the Little Cub's SFR.

The \HI\ flux density is calculated from data collected with the Westerbork Synthesis Radio Telescope (WSRT; \citealt{2001ASPC..240..857B}) and the resulting \HI\ mass is estimated using the equation presented in \citet{2008AJ....136.2563W}. We estimate the stellar mass of the Little Cub using the stellar mass-to-light ratio color correlation given in \citet{2003ApJS..149..289B}, combined with the solar absolute magnitudes reported by \citet{2010MNRAS.404.1215H}.\footnote{Our reported stellar masses have been divided by a factor of 1.26 to correct the "diet" Salpeter IMF employed by \citet{2003ApJS..149..289B} to the \citet{2003PASP..115..763C} IMF, which we use in the SFR.} We report a stellar mass using the $i$-band coefficient in combination with the $r-i$ color of the Little Cub, as these bands are the least affected by the current burst of star formation. We have also removed the contribution of the emission lines from the $r$-band, which amounts to 26\% of the total flux. Based on these calculations, we find that the Little Cub is notably gas-rich, with an \HI\ gas to stellar mass ratio of \textit{M}$_{\textnormal{\HI}}$\,/\,\textit{M}$_{*}$\,$\sim$\,96.


\begin{deluxetable}{ccc}
\tablewidth{0pt}
\tablecaption{Observed and Derived Properties of the Little Cub}
\tablehead{\colhead{Observed Property} & & \colhead{Value}}
\startdata
R.A. (J2000) & & 10$^{\textnormal{h}}$44$^{\textnormal{m}}$42$^{\textnormal{s}}$.66 \\
Decl.  (J2000) & & +63$^{\circ}$06$'$02.08$''$ \\
Redshift & & 0.0032\,$\pm$\,0.0003 \\
$m_{g}$ & & 19.56\,$\pm$\,0.03 \\
$m_{r}$ & & 19.51\,$\pm$\,0.04 \\
$m_{i}$ & & 20.07\,$\pm$\,0.10 \\
\hline \hline
Derived Property & Value & Value \\
\hline
Distance (Mpc) & 15.8 & 20.6 \\
\textit{M}$_{B}$	& $-$11.4 & $-$12.0 \\
\textit{L}(H$\alpha$) (erg s$^{-1}$) & 1.4$\times$10$^{38}$ & 2.5$\times$10$^{38}$ \\
SFR (\textit{M}$_{\odot}$ yr$^{-1}$)	& 0.00063 & 0.0011 \\
\textit{M}$_{\textnormal{\HI}}$ (\textit{M}$_{\odot}$) & 4.7$\times$10$^{7}$ & 8.2$\times$10$^{7}$ \\
\textit{M}$_{*}$ (\textit{M}$_{\odot}$) & 4.9$\times$10$^{5}$ & 8.5$\times$10$^{5}$ \\
Projected Distance to NGC 3359 (kpc) & 69 & 90 \\
\enddata
\tablecomments{Distance estimates to the Little Cub are based on two separate models of the local peculiar velocity flow. We note that all derived properties are dependent on the distance by a factor of $D^{2}$.}
\label{table:properties}
\end{deluxetable}

\subsection{Environment}
The Little Cub has been previously and independently identified as a UV source embedded in an isolated \HI\ cloud near the barred spiral galaxy NGC 3359 \citep{2012MNRAS.426.2441D} and suggested to be a potential satellite of NGC 3359 due to their proximity on the sky ($\sim$14.9$^{\prime}$ separation) and similar heliocentric velocities \citep{1986ApJ...307..453B}. While more precise distance measurements to both NGC 3359 and the Little Cub are required to confirm that the two systems are indeed interacting, the relative velocity of 53~km~s$^{-1}$ suggests that the Little Cub is a satellite of NGC 3359. A recent estimate of the distance to NGC 3359 (20.8~Mpc; \citealt{2013AJ....146...86T}) using the Tully--Fisher Relation is in agreement with our estimate of 20.6~Mpc to the Little Cub using the \citet{2000ApJ...529..786M} model. 
At this distance, the separation between the Little Cub and NGC 3359 on the sky places the Little Cub at a projected distance of 90~kpc from its potential host galaxy.

\begin{figure*}
\includegraphics[width=1.0\textwidth]{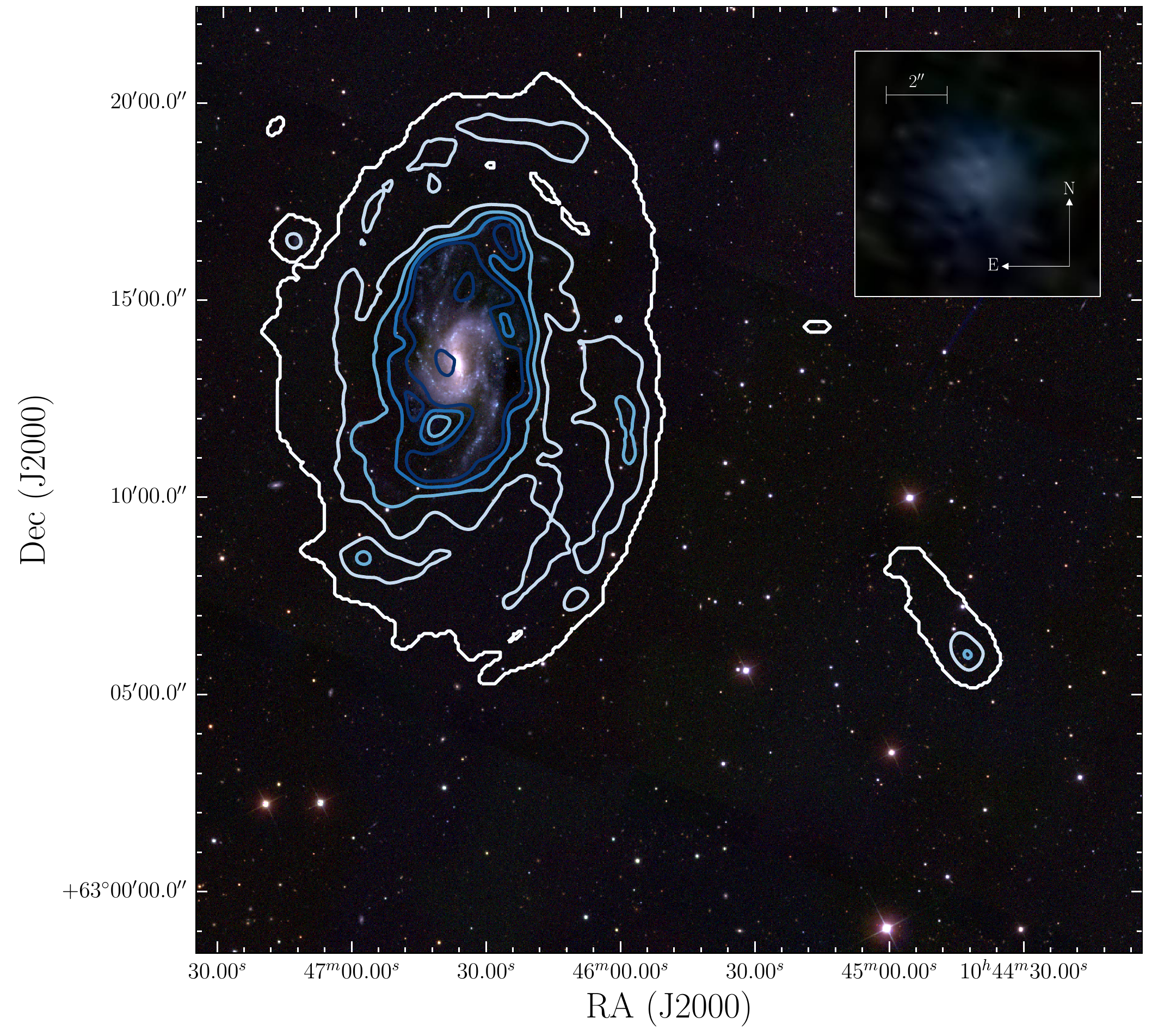}
\caption{Three-color SDSS image of the spiral galaxy NGC 3359 overlaid with \HI\,contours at approximately 0.5, 1.2, 2.4, 3.6, and 4.8$\times$10$^{20}$~cm$^{-2}$ levels, obtained using the Westerbork Synthesis Radio Telescope. The Little Cub is found in the \HI\,tail toward the bottom right of the image, where the \HI\ detection is strongest. A zoom-in of the SDSS image of the Little Cub is shown in the inset at the upper right.}
\label{fig:SDSSWSRT}
\end{figure*}

In Figure \ref{fig:SDSSWSRT}, we show a three-color SDSS image of NGC 3359 and the Little Cub, overlaid with \HI\,contours from WSRT. \HI\ gas is clearly detected around the Little Cub, exhibiting the highest column density in the region of current star formation. We also note an elongation of \HI\ gas in the direction of NGC 3359, which is a strong indication that gas is being stripped from the Little Cub.

In the context of our own Local Group, it is unusual for satellites of more massive galaxies, such as the Little Cub, to contain much gas; all dwarf satellites within 270~kpc of the Milky Way and M31 (with the exception of Leo T and the Magellanic Clouds) are quiescent and undetected in \HI\,\citep{2009ApJ...696..385G}. This is in stark contrast with isolated dwarf galaxies, which are gas-rich and almost always observed to have active star formation.  The environmental differences in which we have found gas-poor and quiescent versus gas-rich and star-forming dwarf galaxies suggest that the timescale for satellite quenching by their massive host is short \citep{2015MNRAS.454.2039F, 2015ApJ...808L..27W}.

However, a recent study by \citet{2017arXiv170506743G} found that the majority of satellites around a sample of eight Milky Way analogs were star-forming, suggesting that Milky Way's satellite population may be atypical. If NGC 3359 and the Little Cub are truly interacting, we may be witnessing a rare example of a low stellar mass dwarf satellite being quenched due to the presence of a more massive host galaxy \citep{2017arXiv170503018S}. The Little Cub will be a particularly intriguing laboratory to test our current understanding of dwarf satellite galaxy evolution, which may be biased by our studies of the Local Group.



\section{Conclusion}
We present Shane+Kast and Keck+LRIS observations of the BCD galaxy J1044+6306, which we nickname the Little Cub, found in the constellation Ursa Major. Our analysis of these spectra show that the Little Cub is one of the lowest-metallicity star-forming galaxies known in the nearby universe, with a direct gas-phase oxygen abundance  of  only12\,+\,log(O/H)\,=\,7.13\,$\pm$\,0.08. We estimate that the Little Cub contains roughly 10$^{5}\,M_{\odot}$ of stars and is gas-rich, with a neutral gas to stellar mass ratio of 96.

We report that the Little Cub exhibits a velocity offset of 53~km~s$^{-1}$ from a nearby grand design spiral galaxy (NGC 3359), at a projected distance of just 69-90~kpc. The Little Cub also shows evidence of neutral gas being stripped, further supporting the idea that these two systems are interacting. While accurate distance measurements to the Little Cub and NGC 3359 are required to confirm their physical proximity, the possible interaction between the two systems provides a unique opportunity to study the contribution of different stripping mechanisms, such as ram-pressure versus tidal stripping, relevant in satellite quenching, as well as the building of more massive galaxies through the accretion of smaller satellite galaxies. 

The Little Cub was selected as a candidate metal-poor system based only on its photometric colors, as part of a larger survey led by the authors to combine SDSS imaging with spectroscopic observations to identify new metal-poor star-forming galaxies in the local universe. To date, this program has yielded highly successful results--we have confirmed about 100 new BCDs, with nearly half the systems estimated to be in the low metallicity regime (T. Hsyu et al. 2017, in preparation), making them less than or equal to a tenth solar metallicity in gas-phase oxygen abundance. This new method is especially promising given the increasing wealth of photometric information that will result from other large area sky surveys such as Pan-STARRS, the Dark Energy Survey (DES), and the Dark Energy Camera Legacy Survey (DECaLS).
\label{Conc}

\acknowledgments
We are grateful to our anonymous referee for their thorough comments that have resulted in an improved manuscript. The authors thank Nissim Kanekar for providing the WSRT data presented in this work, and Marcel Neeleman and Asher Wasserman for helpful guidance in analyzing the data. We are grateful to Kristen McQuinn and Evan Skillman for their expertise and helpful input, and we thank Alis Deason and Connie Rockosi for useful discussions. The data presented herein were obtained at the W. M. Keck Observatory, which is operated as a scientific partnership among the California Institute of Technology, the University of California and the National Aeronautics and Space Administration. The Observatory was made possible by the generous financial support of the W. M. Keck Foundation. The authors wish to recognize and acknowledge the very significant cultural role and reverence that the summit of Mauna Kea has always had within the indigenous Hawaiian community.  We are most fortunate to have the opportunity to conduct observations from this mountain. Research at Lick Observatory is partially supported by a generous gift from Google. We also gratefully acknowledge the support of the staff at Lick and Keck Observatories for their assistance during our observing runs. During this work, R.~J.~C. was supported by a Royal Society University Research Fellowship, and by NASA through Hubble Fellowship grant HST-HF-51338.001-A, awarded by the Space Telescope Science Institute, which is operated by the Association of Universities for Research in Astronomy, Inc., for NASA, under contract NAS5-26555. R.~J.~C. acknowledges support from STFC (ST/L00075X/1). J.~X.~P. acknowledges support from the National Science Foundation grant AST-1412981.

\facilities{Shane (Kast Double spectrograph), Keck:I (LRIS)}
\software{\texttt{astropy} \citep{2013A&A...558A..33A},\, \texttt{matplotlib} \citep{Hunter:2007},\, \texttt{NumPy}\,\citep{2011arXiv1102.1523V},\, \texttt{Seaborn}\,\citep{SeaBorn}}


\end{document}